# The jamming surface of granular matter

# Determined from soil mechanics results

# P. Evesque

Lab MSSMat, UMR 8579 CNRS, Ecole Centrale Paris
92295 CHATENAY-MALABRY, France, e-mail evesque@mssmat.ecp.fr

**Abstract:**

*Classical soil mechanics results are used to propose the equation of the jamming transition surface in the ($\sigma'$, $1/\rho=v$) space, where $\rho$ is the density, $v$ the specific volume and $\sigma'$ the stress really supported by the grain structure. Taking axisymmetric conditions, labelling $q=\sigma'_{11}-\sigma'_{22}$ and $p'=(\sigma'_{11}+\sigma'_{22}+\sigma'_{33})/3$, and considering normal range of pressure (10 kPa-10MPa) the equation of the surface of jamming transition is $v=v_o-\lambda \ln(p'/p'_o)+ \lambda_d \ln(1+q^2/(M'p')^2)$; M' is related to the friction angle $\varphi$, $\lambda$ and $\lambda_d$ are two constants which depend on soil characteristics.*

______________________________________________________________________

Jamming transition is a fundamental problem which attracts some interest from physicists recently, since they have addressed a parallel with the glass transition [1]. We do not want to discuss this last point here; but we just want to stress that some information on the jamming transition can be found already from the technical literature, within different words most likely; for instance, the soil mechanics literature speaks of this problem within an other terminology, but it has already identified the law of variation of the minimum density that a soil can exhibit in statics and under a definite stress field. This may help physicists in their investigation. This is just what we want to report.

Consider a dry granular material (as sand) in static condition, submitted to an axial stress field expressed in the principal axis direction as ($\sigma'_{11}$, $\sigma'_{22}$, $\sigma'_{33}$), with $\sigma'_{11} > \sigma'_{22}=\sigma'_{33}$; it can be built at different density $\rho=1/v$. However, this density cannot be looser than a given value; this loosest state is called the "normally consolidated state" in the mechanics literature [2,3,4]; it is characterised by its specific volume $v_{nc}$. $v_{nc}$ is found to depend on the stress field. Labelling $q=\sigma'_{11}-\sigma'_{22}$, $p=(\sigma'_{11}+\sigma'_{22}+\sigma'_{33})/3$, $\eta=q/p'$ and M' the ratio q/p' at the limit of plasticity, i.e. $M'=6\sin\varphi/(3-\sin\varphi)$ with $\varphi$ the friction angle, one gets the following equation for $v_{nc}$ from experimental fit :

$$v_{nc}=v_{nco}-\lambda \ln(p'/p'_o) - \lambda_d \ln(1+\eta^2/M'^2) \qquad (1)$$

where $\lambda$ and $\lambda_d$ are two constants which depend on material. The domain of validity of this equation is (10 kPa, 10 MPa). For smaller pressure range, experiments in micro-gravity experiments have to be performed and are currently being performed by NASA; it is probable that $v_{nc}$ tends to a given limit. For pressure larger than 10 MPa, grain crushing occurs, modifying the $v_{nc}$ law of variation; examples can be found in [3]. λ is about 0.06 granular matter and sands; an estimate of this value has been tentatively proposed from a microscopic modelling [4,5].

When granular matter is saturated with liquid, it is found experimentally that Eq. (1) still holds, but the stress field which has effectively to be considered in Eq. (1) is the effective stress, i.e. the one which is really carried by the grain structure. So owing to the so-called Terzaghi approximation, σ' is then equal to the total stress $\sigma_{tot}$ minus the liquid pressure $u_w$:

$$\sigma' = \sigma_{tot} - u_w \qquad (2)$$

It is found that λ and $\lambda_d$ does not depend on the presence of saturating liquid.

When considering clays saturated with water, similar results are still valid, and Eq. (1) holds, if Eq. (2) is taken into account. The main difference is that the values of λ and $\lambda_d$ are a bit larger: the material is more "compressible" indeed. The name "normally consolidated" comes from clays since it is much more complicated to get loose sand samples than clay ones. Examples of behaviour of clays can be found in [3].

It is worth mentioning that the shape of the transition curve in the (v=1/ρ, σ') space is convex instead of concave as proposed in ref. [1].

As a final remark, the ensemble of normally consolidated states forms what is called the Roscoe's surface in soil mechanics literature when q/p'=η<M'; It forms what is called the Hvorslev's surface when q/p'=η>M'. Both surfaces are parts of the same surface [5,6]. Knowing these notations may help physicists in finding more information.

*Acknowledgements:* CNES is thanked for partial funding.

The electronic arXiv.org version of this paper has been settled during a stay at the Kavli Institute of Theoretical Physics of the University of California at Santa Barbara (KITP-UCSB), in june 2005, supported in part by the National Science Fundation under Grant n° PHY99-07949.

*Poudres & Grains* can be found at :
http://www.mssmat.ecp.fr/rubrique.php3?id_rubrique=402